\documentclass[useAMS,usenatbib,usegraphicx]{raa}
\usepackage{graphicx}
\usepackage{natbib}

\usepackage{amsmath}    % Advanced maths commands
\usepackage{multirow}
\usepackage{setspace}
\begin{document}

\title{Outflows and Young Stars in the Auriga-California Giant Molecular Cloud}

 \volnopage{Vol.0 (202x) No.0, 000--000}  %%preserved for Editor. DOn't remove!  \setcounter{page}{1}     %%starting page, preserved for Editor. DOn't remove!    
\author{T.A. Movsessian,\inst{} T.Yu. Magakian\inst{}} 
%\footnotemark[1]\thanks{This work is  based on observations conducted with the 1-m Schmidt telescope of the Byurakan Astrophysical Observatory of the National Academy of Sciences of Armenia.}\\
 \institute{Byurakan Observatory NAS Armenia, Byurakan, Aragatsotn prov., 0213, Armenia  {\it tigmov@bao.sci.am; tigmag@sci.am}\\
}

\date{}

\vs\no                                                                             {\small Received 202x month day; accepted 202x month day} 

\abstract 
{The Auriga-California Molecular Cloud (AMC) is a nearby star-forming region with a moderate level of star formation activity. Previous infrared surveys have shown that its young stellar population is concentrated along a dense filament extending north of NGC 1579. We searched for new manifestations of ongoing star formation in the AMC by identifying optical and molecular outflows, H$\alpha$ emission-line stars, and young stars associated with reflection nebulae. Narrow-band H$\alpha$ and [\ion{S}{ii}] imaging, together with slitless spectroscopy, were obtained with the 1-m Schmidt telescope of the Byurakan Observatory. The optical data were complemented by near-infrared images from the \textit{Spitzer} Enhanced Imaging Products (SEIP) and by WISE/unWISE archival data. We discovered seven new Herbig-Haro (HH) objects and outflows associated with several young stellar objects, including deeply embedded infrared sources and stars surrounded by compact reflection nebulae. Nearly all of the optical HH objects have molecular hydrogen counterparts, and several additional candidate H$_2$ outflows without optical counterparts were identified. Particularly noteworthy is a compact region around V347 Per that appears to host at least three molecular outflows. We also present a list of newly identified H$\alpha$ emission-line stars and a catalogue of stars associated with optical reflection nebulae. Both populations are concentrated primarily along the northern dark filament of the cloud. The spatial distribution of the newly identified HH objects, molecular outflows, emission-line stars, and nebulous young stars confirms that the northern filament is the principal site of ongoing star formation in the AMC. The close correspondence between the optical and molecular outflows further suggests that most of their driving sources are deeply embedded Class 0/I young stellar objects.
}
  \titlerunning{Outflows and young stars in AMC}

  \maketitle

\section{Introduction}
It is now widely accepted that young stellar objects (YSOs) undergo a phase of mass outflow, which plays an essential role in the process of star formation  \citep{Arce,BRD}. 

At optical wavelengths, manifestations of mass outflow are observed as Herbig-Haro (HH) objects, which are produced either by the interaction of faster gas clumps with slower material ejected earlier or by the interaction of supersonic gas ejected from YSOs with the surrounding interstellar medium  \citep{RB}.
Therefore, HH objects are excellent tracers of recent star formation. The identification of HH objects and molecular hydrogen objects (MHOs) can also provide clues to the locations of protostars, especially those deeply embedded within molecular cloud cores. 

To increase the number of known outflows in nearby star-forming regions, we initiated several years ago a survey using the 1-m Schmidt telescope at the Byurakan Astrophysical Observatory, the Byurakan Narrow Band Imaging Survey (BNBIS). In this paper, we present the results of our search in the Auriga-California Molecular Cloud (AMC).

According to recent measurements, the AMC is located at a distance of $532 \pm 28$ pc \citep{Omodaka2020}. Despite its overall similarity to the Orion A molecular cloud, the AMC does not produce highly luminous OB stars. Its most prominent young stellar object is the nebulous source LkH$\alpha$ 101,
a probable  HAeBe emission-line star. In the infrared\textbf{}, the heavily reddened LkH$\alpha$ 101 is one of the brightest
young stellar objects in the sky. Located just north of the Taurus-Auriga complex in the L1482
dark cloud, it appears to be an early B-type star, although no stellar absorption lines have yet been detected, and its circumstellar disk has been directly-imaged \citep{Tuthill}. The AMC has been the subject of numerous observational studies (see the review by \citealt{Andrews_Wolk}). In addition to LkH$\alpha$ 101, it   contains a group of
H$\alpha$ emission-line stars and at least 5 B-type stars
\citep{H-A-D}.

As shown by mid- and far-infrared studies \citep{Harvey2013,Broekhoven}, a dark filament extends for approximately one degree north of  LkH$\alpha$ 101.  Along this filament, these authors identified a chain of infrared sources representing young stellar objects. At optical wavelengths (e.g., in Pan-STARRS images), several stars associated with small reflection nebulae and dark dust filaments can also be seen.    However, this part of AMC has never been investigated in detail in the optical range.  It was in this region that we discovered a new HH outflow, HH 1228  \citep{Mov2024}. In the present paper, we continue our study of the AMC by combining narrow-band imaging with objective-prism slitless observations. We report the discovery of new HH knots and outflows, emission-line stars, and newly identified molecular hydrogen objects.

\section{Observations and Data Reduction}

The observations were carried out between 2020 and 2024 with the 1-m Schmidt telescope of the Byurakan Astrophysical Observatory. The detector was a liquid-cooled  4K $\times$ 4K Apogee (USA)  CCD camera, which a pixel scale of 0.868\arcsec\ and a field of view of approximately 1$^\circ$\  \citep{Dodo}. Narrow-band filters
centered on 6560 \AA\ and 6760 \AA, both with a full width at half maximum (FWHM) of 100 \AA, were
used to obtain H$\alpha$ and [\ion{S}{ii}] images, respectively. A
medium-band filter, centered on 7500 \AA\ with a FWHM of 250 \AA, was
used for continuum imaging. A dithered set of 5 min exposures was obtained in each
filter.  The continuum exposures were obtained immediately after the narrow-band observations, with 1800 s of total integration time. We estimated the limit for our images taken with narrow-band filters to be 21$^{m}$ for point sources (stars) in the \textit{r}  band from PanSTARRS survey and 5400 s of integration time for point sources (stars).

In addition, slitless observations were performed using a 4-degree objective prism. The dispersion of this prism near  H$\alpha$ is 2.5 \AA/pix, but effective resolution is defined mainly by seeing (about 6-7 \AA\ or R$\approx$\ 1000 with 2\arcsec\ seeing).   To avoid spectral overlap, a medium-band filter centered on the H$\alpha$ line with a bandwidth of approximately 200 \AA\ was employed. The fields, filters, total integration times, and the dates of observations are listed in Table \ref{log}. 

 The images were reduced following standard procedures  using an IDL package developed by
S.~Dodonov (SAO RAS), including bias subtraction, cosmic-ray
removal, and flat-field correction using a superflat constructed from
a set of science images.

The search for HH objects was performed using the classical technique,
proposed by \citet{VDB}, based on comparison of the H$\alpha$,
[\ion{S}{ii}] and continuum images. Over the years, this method has proven
to be highly reliable for identifying HH objects in the vast majority of
cases.
Consequently, narrow-band imaging in [\ion{S}{ii}] combined with adjacent continuum observations provides an efficient and reliable approach for surveying star-forming regions for HH objects.

The observations covered three fields in the central part of the Auriga-California molecular cloud, as shown in Fig.\ref{field}. The coordinates of their centers are the following: ``North'': RA = 04$^{h}$ 30$^{m}$ 37$^{s}$, Dec = +35\degr\ 55\degr\ 57\arcsec\ (2000.0), ``Center'': 04$^{h}$ 33$^{m}$ 30$^{s}$ +35\degr\ 20\arcmin\ 34\arcsec\ , ``South'': 04$^{h}$ 30$^{m}$ 58$^{s}$ +35\degr\ 01\arcmin\ 25\arcsec\ correspondingly.

To search for candidate molecular hydrogen objects (MHOs), we used near-infrared images from the \textit{Spitzer} Enhanced Imaging Products (SEIP), which provide improved spatial resolution. We concentrated on data obtained with the Infrared Array Camera (IRAC), particularly the 4.5~$\mu$m band, which is well known to trace molecular outflows from young stellar objects \citep{noriega2012}. Although these broadband mid-infrared observations cannot replace narrow-band imaging in the 2.12~$\mu$m H$_2$ line, they can reveal candidate H$_2$ outflows suitable for subsequent detailed follow-up observations.

%As in the optical range, we use emission lines NIR images help identify %outflow systems, which consist of Molecular Hydrogen Objects (MHOs) of shock %excitation.

To identify bright infrared sources and additional MHO candidates, we also used data from the WISE survey, comparing the unWISE W1 and W2 images, since these bands closely correspond to the IRAC 1 and IRAC 2 bands, respectively. Several authors have demonstrated that extended excess emission at 4.5~$\mu$m is an effective indicator of molecular outflows in the mid-infrared \citep{Cy2008,Chambers2009,Zhang2014}.

\begin{table}
\begin{center}
\caption[]{Log of 1-m Schmidt Observations}       
\label{log}     
{\setstretch{1.3}        
\begin{tabular}{l c c }          
\hline\noalign{\smallskip}                        
Field and filter & Total Exposure time (s) & Obs. date  \\    
\hline\noalign{\smallskip}                     
Center  H$\alpha$ &6000 &2020 Oct 13 \\
Center [SII] & 7200 & 2020 Oct 15 \\
North  H$\alpha$ & 4800 & 2022 Nov 24 \\ 
North [SII] & 7200 & 2023 Jan 10 \\
South   H$\alpha$ & 1800& 2023 Jan 23 \\ 
South   H$\alpha$ & 3500 & 2023 Jan 24 \\
South [SII] & 6000 & 2024 Nov 24 \\
North  H$\alpha$ (slitless)  & 6300 & 2024 Dec 03 \\   
South   H$\alpha$ (slitless) & 3600 & 2024 Dec 03 \\

\noalign{\smallskip}\hline                                       
\end{tabular}
}
\end{center}
\end{table}

\section[]{Results}
\subsection{Overview\\ }

As was described above, our observations covered three fields tracing a star-forming filament extending in the north-south direction and containing numerous YSOs \citep{Harvey2013}. Prior to our study, no evidence of outflow activity had been reported in this region, either in CO or at optical wavelengths, until the discovery of a chain of HH objects (HH 1228) not far from LkH$\alpha$ 101. This outflow, clearly visible in the \textit{Spitzer} SEIP mid-infrared survey, has a projected length of about 1 pc \citep{Mov2024}.

The first field was centered on the young stellar object LkH$\alpha$ 101 (hereafter the ``central field''). The second field covered the region approximately $1^\circ$ north of LkH$\alpha$ 101, which contains several far-infrared sources as well as a number of nebulous stellar objects (hereafter the ``northern field''). The third field (the ``southern field'') was centered on the group of \textit{Herschel} sources located about 30\arcmin\ south of LkH$\alpha$ 101 (see Fig.~8 in \citealt{Harvey2013}).

\begin{figure}
  \centering
   \includegraphics[width=250pt]{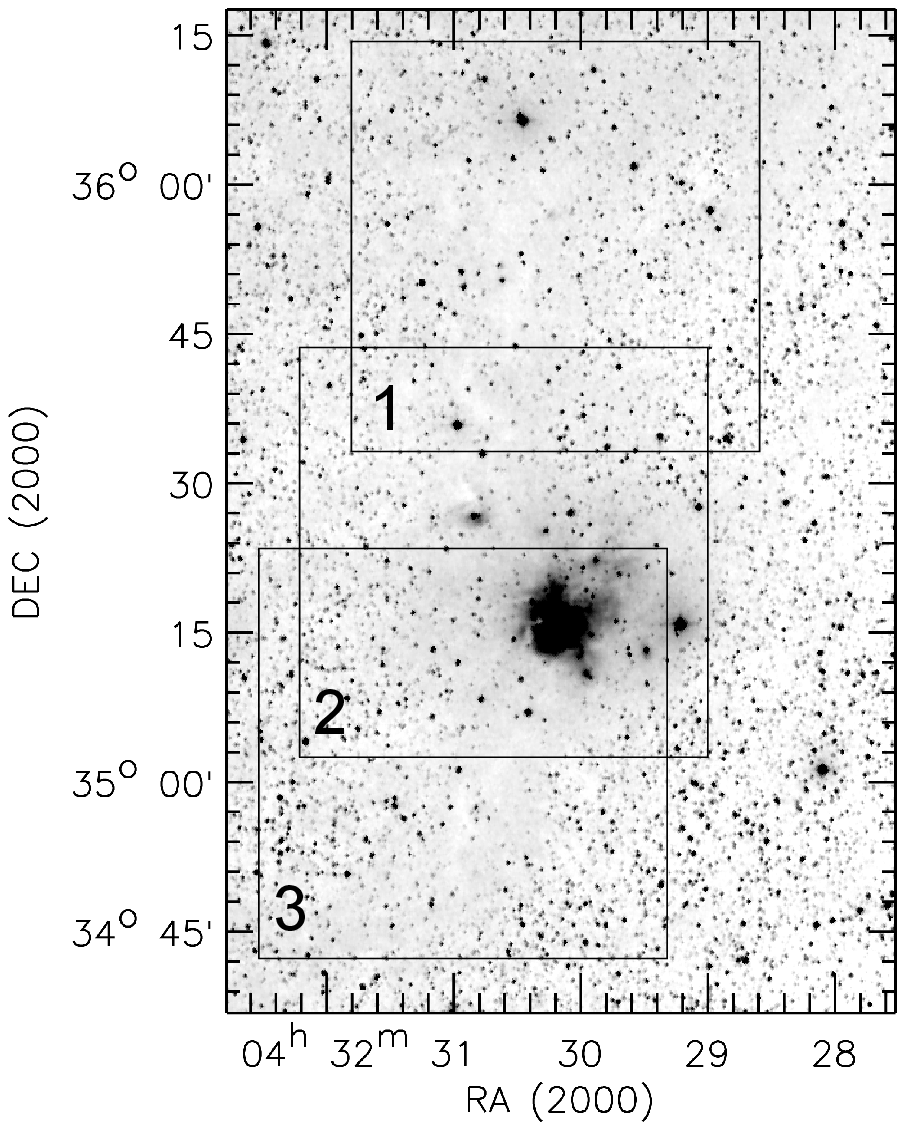}
     \caption{DSS2-R image of  Auriga-California molecular cloud with superposed position of the fields obtained with the 1 m Byurakan Schmidt telescope. The bright nebulous object near the center is LkH$\alpha$ 101 young star with NGC 1579 reflection nebula. Numbers 1,2,3 correspond  to  northern, central and southern fields.}
   \label{field}
\end{figure}

  Figure~\ref{field} shows the DSS2-R image of the AMC with the observed fields marked by rectangles. In the following sections, we discuss each field separately. The newly discovered HH objects and the candidate molecular outflows are listed in Tables~\ref{HHos} and \ref{MHOs} and are shown in Figs.~2--7. Since final MHO numbers can be assigned only to flows confirmed by observations in 2.12~$\mu$m H$_2$ line, we use a preliminary numbering scheme in the form of \textit{m.n}, where \textit{m} is the number of field and \textit{n} is the number of candidate molecular flow in this field. For convenience, we list the coordinates in the table \ref{MHOs} with a precision of 0.1$^{s}$ in RA, although the actual measurement accuracy based on the SEIP images is somewhat lower.

\subsection{Northern field}
% Our observations have revealed several HH knots, which are mainly associated %with deeply embedded YSOs. On the Fig.\ref{N_field} the images of the part %of northern field in H$\alpha$ plus [SII], obtained with Schmidt telescope, %are shown. Newly discovered HH objects are marked by arrows. They are described %in detail below.

%\begin{figure}
%  \centering
%  \includegraphics[width=250]{N_smt.eps}
%   \caption{H$\alpha$ + [SII] image of the northern region obtained with %the Schmidt telescope. Newly discovered HH objects and emission stars are %marked by ellipses and arrows respectively.}
%   \label{N_field}
%\end{figure}

\subsubsection{Area around 2MASS J04303932+3552006 and 2MASS J04303916+3552038}

These two YSOs are located nearby (projected separation is 4\arcsec), near the apex of a bright cone-shaped reflection nebula. Around 2MASS J04303932+3552006, faint wisps of the nebula form a small ring-like structure. These details are clearly visible in the Pan-STARRS images. The entire group is embedded within the dark filament.

Approximately 1.5\arcmin\ to the south-southwest, another group of nebulous optical/infrared sources is located, the brightest of which is 2MASS J04303756+3550320. About 35\arcsec\ south-southeast of this object, the Pan-STARRS images reveal a faint elongated nebulous source coincident with 2MASS J04303831+3549591. This source is barely detectable both in the optical images and in the 2MASS and WISE surveys.

\begin{figure*}
  \centering
    \includegraphics[width=450pt]{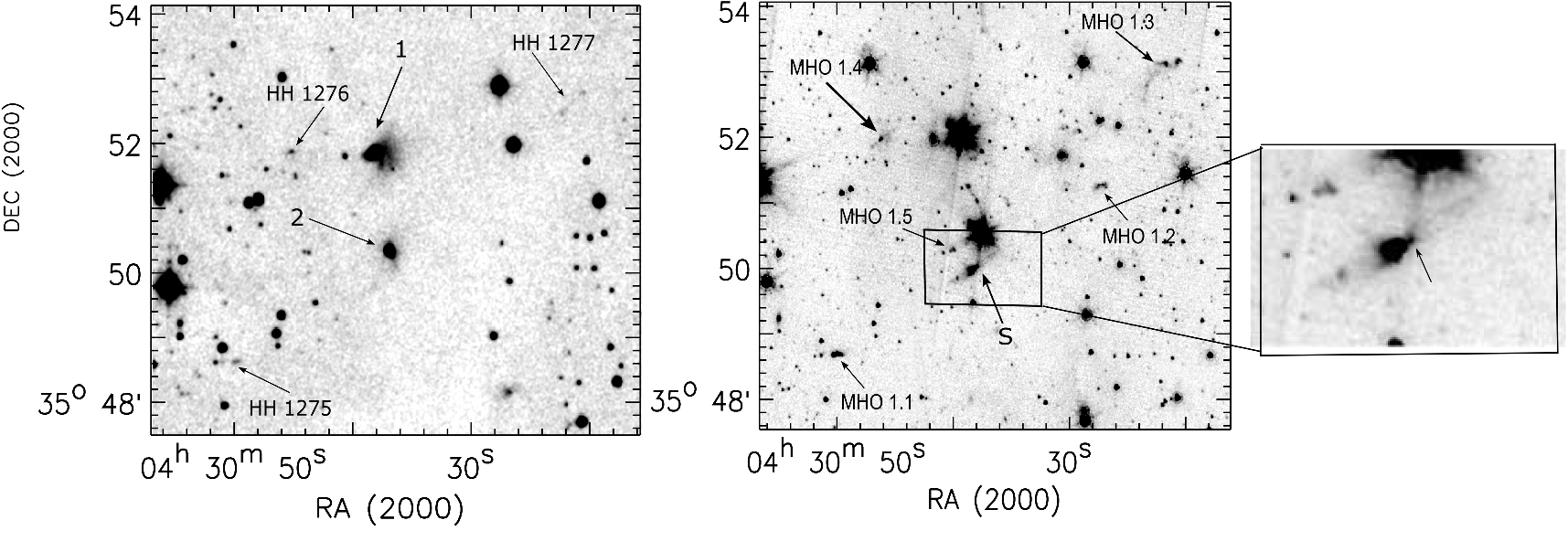}
   \caption{ Left: H$\alpha$ + [SII] image of the area with HH 1275, HH 1276 and HH1277. By numbers 1 and 2 the  pair of stars 2MASS J04303932+3552006 \&\ 2MASS J04303916+3552038 and 2MASS 04303756+3550320 are marked correspondingly. Right: the same area in  4.5 $\mu$m. Five molecular hydrogen  knots are marked. By letter ``S" the embedded source 2MASS J04303831+3549591 is denoted. The field around it is shown also enlarged.  }
      \label{HH1}
\end{figure*}

Our narrow-band observations revealed three faint HH objects distributed across this area. They are indicated by arrows in the left panel of Fig.~\ref{HH1}, which presents the combined H$\alpha$ + [\ion{S}{ii}] image. The right panel shows the SEIP IRAC2 image, in which five groups of H$_2$ emission knots can be identified. Three of them coincide precisely with the newly detected HH objects.

The northernmost HH object, HH 1277, consists of two stellar-like knots A and B, connected by a faint emission filament. It is located about 4\arcmin\ northwest of the nebulous source 2MASS J04303756+3550320. The emission structure is elongated approximately toward this probable driving source. In the SEIP image, HH 1277 has a molecular counterpart (MHO 1.3) in the form of a slightly arcuate flow oriented toward the northeast. It contains at least three bright knots, two of which coincide exactly with the optical knots of HH 1277.

The second HH object (HH 1276) appears as a compact emission knot located approximately 1.4\arcmin\ east of 2MASS J04303932+3552006. It also has an H$_2$ counterpart (MHO 1.4).

The third object (HH 1275) lies about 3\arcmin\ southeast of the nebulous source 2MASS J04303831+3549591 and appears as an elongated emission knot. It coincides remarkably well with the H$_2$ condensation MHO 1.1, which has the same morphology and dimensions.

In addition, a clearly defined bipolar molecular jet associated with 2MASS J04303831+3549591 is visible (Fig.~\ref{HH1}, right). We suggest that this embedded infrared source is the most probable driving source of molecular knots 1.1 (associated with HH 1275) and 1.2 (which has no optical counterpart), since both knots lie on a common axis passing through the source and are located at nearly equal projected distances from it. HH 1277 is situated close to the extension of the same axis and may also belong to the western lobe of this outflow. Alternatively, HH 1277 and MHO 1.3 may be driven by one of the nearby infrared nebulous stars. HH 1276 together with MHO 1.4 may originate from one of the stars from the pair 2MASS J04303932+3552006 and 2MASS J04303916+3552038. MHO 1.5 is most likely associated with 2MASS J04303756+3550320.

\subsubsection{Area around IRAS 04269+3550}

\begin{figure*}
  \centering
  \includegraphics[width=450pt]{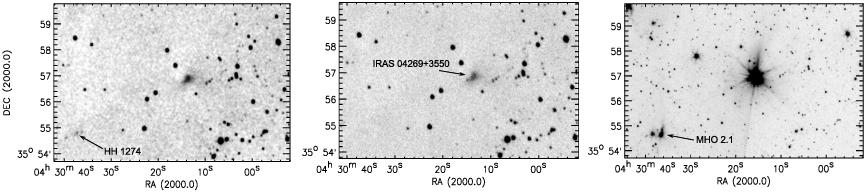}
   \caption{ Left: H$\alpha$ + [SII] image of the HH 1274 flow. Three knots inside HH 1274 are well distinguishable. Center: the same area in continuum; IRAS 04269+3550 is marked. Right: the same area in  4.5 $\mu$m. Molecular hydrogen knot is marked. }
   \label{HH2}
\end{figure*}

IRAS 04269+3550 is barely visible in the PanSTARRS \textit{i-y} bands as a stellar-like source near the apex of a very faint cone-shaped reflection nebula. In near-infrared 2MASS images, however, it  appears relatively bright, and the nebula is seen as a well-defined broad cone opening toward the northwest.  In the WISE images, this source becomes the brightest object in this field. 
Our narrow-band observations revealed a chain of three emission knots (HH 1274 A, B, and C) located about 5\arcmin\ southeast of IRAS 04269+3550 (Fig.~\ref{HH2}). The knots lie along a common axis passing through the infrared source, on the side opposite to the reflection nebula. The projected length of the chain is about 40\arcsec. The SEIP IRAC2 image shows a bright molecular counterpart to this group. Interestingly, this probable MHO 2.1 is elongated in a direction nearly perpendicular to the axis defined by the optical knots (Fig.~\ref{HH2}, right).
We assume that such a configuration is a not unlikely intersection of two flows, but instead in the mid-IR we observe a bright bow-shock of the same flow (note that the apex of this arc is oriented in the same direction as the optical flow).

\subsubsection{Area around 2MASS J04302467+3545204}

Closer to the southern edge of this field, the Pan-STARRS images reveal a pair of nebulous stars. One of them, 2MASS J04302705+3545505, illuminates a compact bipolar reflection nebula with broad opposing lobes, while the second source, 2MASS J04302615+3545578, is also surrounded by faint nebulosity that is particularly evident in the HST images. However, both objects are relatively faint at infrared wavelengths.

\begin{figure*}
  \centering
  \includegraphics[width=450pt]{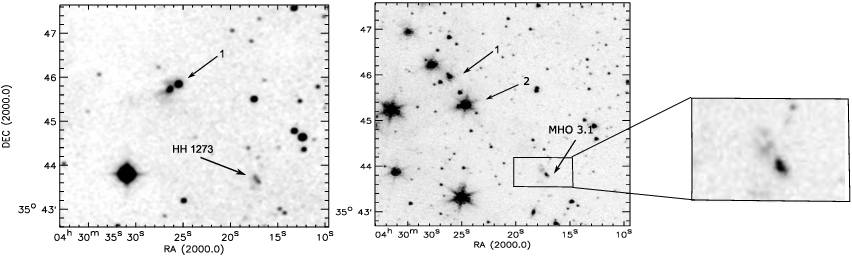}
   \caption{ Left: H$\alpha$ + [SII] image of the HH 1273 flow. Right: the same area in  4.5 $\mu$m. The pair of nebulous stars 2MASS J04302705+3545505 and 2MASS J04302615+3545578 is marked as number 1. Note their faintness in the IR frame. The probable source of the flow 2MASS J04302467+3545204 is marked as number 2. The area of MHO 3.1 is shown enlarged.}
   \label{HH1273}
\end{figure*}

Approximately 40\arcsec\ south of this pair lies 2MASS J04302467+3545204, a deeply embedded infrared source that is invisible at optical wavelengths and remains undetected even in the Pan-STARRS \textit{y} band. Based on its infrared colours, this object was classified as a candidate YSO \citep{Dunham}. About 1\arcmin\ southwest of 2MASS J04302467+3545204, we discovered a relatively bright, elongated spearhead-shaped HH knot (HH 1273) with a projected length of about 25\arcsec, oriented away from the infrared source (Fig.~\ref{HH1273}). The knot has a bright probable molecular counterpart 3.1 whose morphology closely resembles a narrow bow shock with the same orientation (Fig.~\ref{HH1273}, right). We therefore suggest that 2MASS J04302467+3545204 is the most probable driving source of this outflow.

\begin{figure*}
  \centering
  \includegraphics[width=450pt]{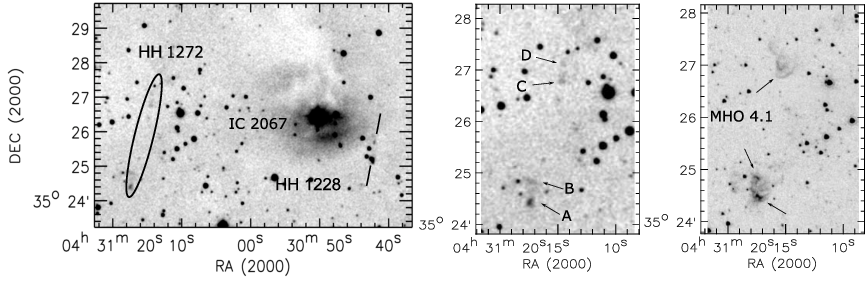}
   \caption{ Left: H$\alpha$ + [SII] image of the HH 1272 flow (a chain of knots inside the ellipse). The previously found HH 1228 flow is shown by arrows near the western edge of the frame. Center: the enlarged field of the HH 1272 from the same image. Main knots are denoted by letters. Right: the same area in 4.5 $\mu$m. The arcuate structures of the probable MHO flow  (4.1) are pointed by  arrows.   The probable source IRAS 04275+3531 is outside of the upper border of this frame.  }
   \label{center}
\end{figure*}

\begin{figure*}
  \centering
  \includegraphics[width=500pt]{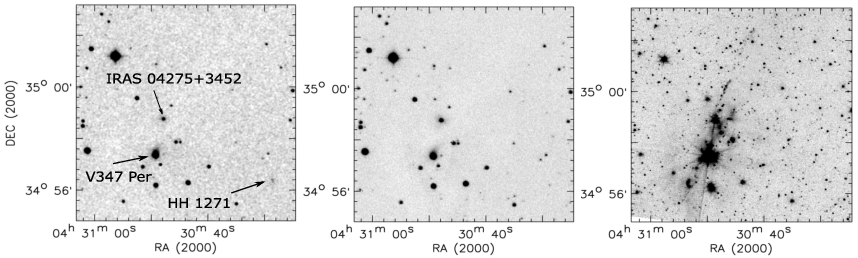}
   \caption{H$\alpha$ + [SII] image of the HH 1271, located near V347 Per (left); the same area in continuum (center); the same area in 4.5 $\mu$m  (right).  }
   \label{south}
\end{figure*}

\subsection{Central field}

We previously discovered an extended outflow system consisting of a chain of five HH knots (HH 1228) east of the nebulous star Gaia DR3 173369863892701312, located near the centre of this field \citep[see Sect.~3.2.3 and the identification chart therein]{Mov2024}. This system has a molecular hydrogen counterpart that is clearly visible in the SEIP IRAC2 images. Its most probable driving source is the very red object WISE J043041.16+352940.7, which is associated with a well-defined conical infrared reflection nebula whose symmetry axis is aligned with the HH 1228 outflow (see Fig.~4 in \citealt{Mov2024}).

A further examination of this region revealed a faint clumpy emission structure located about 6\arcmin\ east of the same Gaia DR3 173369863892701312 star. Its molecular counterpart 4.1 is much more prominent in the SEIP images, where bow-shaped structures with their apices pointing south coincide with the positions of the newly discovered HH knots (Fig.~\ref{center}). The driving source of this outflow is not obvious. No suitable candidate is visible either in the Pan-STARRS or in the 2MASS images. The only possible YSO is IRAS 04275+3531, which lies along the axis defined by the HH knots. However, its projected separation from the outflow exceeds 14\arcmin. Nevertheless, this source is associated with a small fan-shaped infrared reflection nebula opening toward the south, as reported by \citet{CRT2007}.

We conclude that the central part of the dark filament contains two extended outflow systems whose driving sources cannot yet be identified with confidence.

\subsection{Southern field} 

Approximately 20\arcmin\ southwest of LkH$\alpha$ 101, within the southern extension of the dark filament, lies the relatively bright nebulous young stellar object V347 Per. It is located near the centre of a group of Herschel far-infrared sources (Fig.~\ref{south}), only two or three of which are detected at optical wavelengths. Since its discovery as a variable star in 1965 \citep{maffei}, V347 Per has received relatively little attention. Only recently has it been included in optical and infrared surveys of the Auriga-California Molecular Cloud.

Our narrow-band observations revealed only a single faint emission knot, located about 4\arcmin\ west of V347 Per. This object is designated HH 1271 (note that it lies almost coincident with a faint field star, although the two objects are separated by several arcseconds). Its association with V347 Per remains uncertain.

\begin{figure}
  \centering
  \includegraphics[width=200pt]{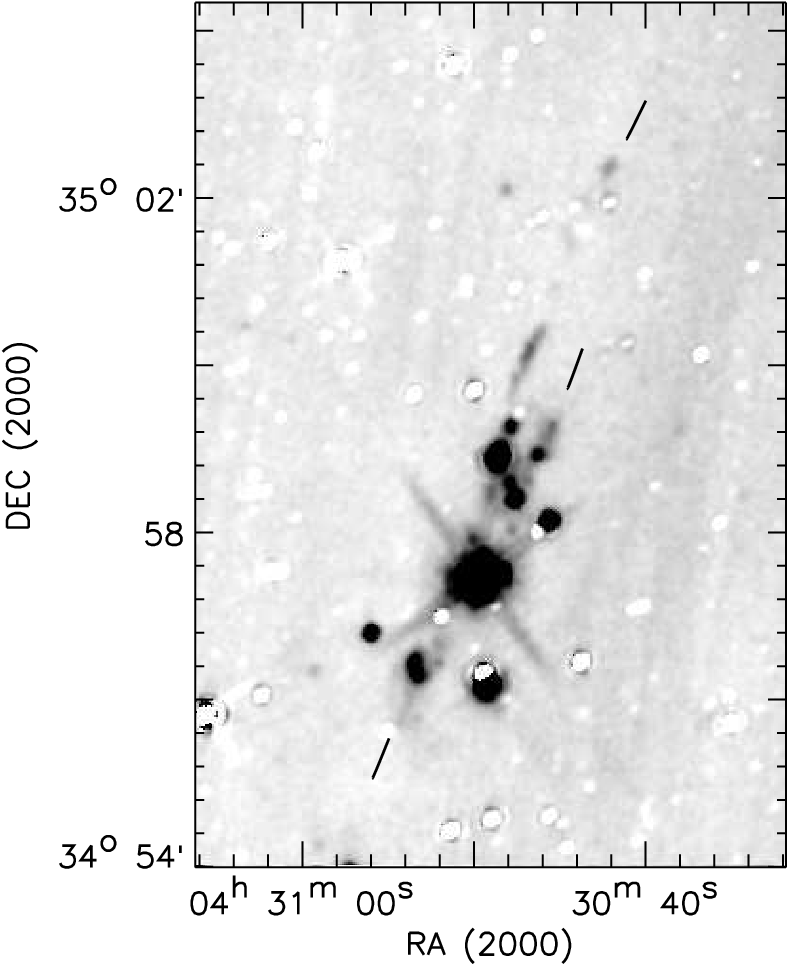}
   \caption{The difference between unWISE W2 and W1 images of V 347 Per and its surroundings. Probable molecular hydrogen flows are marked with dashes.   }
   \label{w2-w1}
\end{figure}

However, an examination of the compact group surrounding V347 Per in the unWISE W2 images, as well as in the \textit{Spitzer} IRAC2 data, revealed several jet-like structures. To enhance the visibility of the extended emission, Fig.~\ref{w2-w1} presents the difference image obtained by subtracting the unWISE W1 image from the W2 image.

 The infrared morphology clearly suggests that this group constitutes a compact cluster of active outflow sources. At least three narrow probable H$_2$ outflows (preliminary designated as 5.1, 5.2 and 5.3) can be identified in Fig.~\ref{w2-w1}. Although the angular resolution is insufficient to identify their driving sources unambiguously, we suggest that outflows 5.1 and 5.2 represent the jet and counterjet of a single source, since they are aligned along a common axis. Outflow 5.3 may instead originate from V347 Per itself.

Taking into account the 530 pc as the probable distance of the cloud, we can conclude, that the majority of the HH flows do not exceed 1 pc in the full length, with the only probable exception
 of HH 1272, which projected distance from the source can be more than 2 pc.

\begin{table*}
\caption[]{List of newly discovered HH objects}       
\label{HHos}     
\setlength{\tabcolsep}{2pt}
\small
\centering
{\setstretch{1.4}}        
\begin{tabular}{l c c c c}          
\hline\noalign{\smallskip}                        
 HH knot & RA (2000.0) & Dec (2000.0) & Probable source & r\arcsec\ $^{a}$  \\
   & \ \ h \ m \ s & \ \ \degr\ \ \arcmin\ \ \arcsec\ \\    
\hline\noalign{\smallskip}                     
 
North  \\
    
 HH 1277A & 04 30 23.3 & +35 52 49 & \\    
 HH 1277B & 04 30 21.9 & +35 53 07 & \\   
 HH 1276 & 04 30 46.2 & +35 51 59 & 2MASS J04303932+3552006 & \\   
 HH 1275 & 04 30 50.1 & +35 48 42 & 2MASS J04303831+3549591 & 174 \\

 HH 1274A & 04 30 40.0 & +35 54 35 & IRAS 04269+3550 & 331 \\
 HH 1274B & 04 30 37.6 & +35 54 48 & \\   
 HH 1274C & 04 30 36.5 & +35 54 50 & \\    
 HH 1273 & 04 30 17.3 & +35 43 52 & 2MASS J04302467+3545204 & 130  \\  
 Center\\ 
 
 HH 1272A & 04 31 17.8 & +35 24 27 & IRAS 04275+3531? &886 \\
 HH 1272B & 04 31 17.8 & +35 24 50 & \\   
 HH 1272C & 04 31 15.3 & +35 26 51 & \\   
 HH 1272D & 04 31 15.2 & +35 27 21 & \\    
 South\\
 HH 1271  & 04 30 28.9 & +34 56 42 &  \\

\noalign{\smallskip}\hline                                       
\end{tabular}
\begin{flushleft}
\textit{Notes}: \\
$^{(a)}$  The projected distance from the probable source (only if unambiguously identified)\\

\end{flushleft}

\end{table*}

\begin{table*}
\caption[]{List of newly discovered candidate MHO}       
\label{MHOs}
\setlength{\tabcolsep}{1pt}     
\centering
\small
{\setstretch{1.3}}        
\begin{tabular}{c c c c c p{1.5cm} }          
\hline\noalign{\smallskip}                        
Field and MHO cand.  & RA (2000.0) & Dec (2000.0) &Probable source & Opt. counterpart & Comments \\
    & \ \ h \ m \ s & \ \ \degr\ \ \arcmin\ \ \arcsec\ \\    
\hline\noalign{\smallskip}                     
1 (2MASS J04303932+3552006) \\ 
 1.1 & 4 30 50.1 & +35 48 42 & 2MASS J04303831+3549591? & HH1275& elongated  \\
 1.2 & 4 30 27.3 & +35 51 18 & 2MASS J04303831+3549591? & no & elongated, 2 knots  \\
 1.3 & 4 30 20.5 & +35 53 12 & &HH1277  & 3 knots in a line$^a$  \\
 1.4 & 4 30 46.2 & +35 51 59 & 2MASS J04303932+3552006? & HH 1276 \\
 1.5 & 4 30 40.1 & +35 50 17 & 2MASS J04303756+3550320 & no\\
2 (IRAS 04269+3550)\\
 2.1 & 4 30 36.7 & +35 54 52 & IRAS 04269+3550 & HH 1274 & arcuate\\
3 (2MASS J04302705+3545505)\\
3.1 & 4 30 17.1 & +35 43 49 & 2MASS J04302467+3545204 & HH 1273 & spear-like$^a$\\
4 (Center)\\
4.1 & 4 31 17.6 & +35 24 30 & IRAS 04275+3531? & HH 1272 & extended for at least 3\arcmin, several bow-shocks$^a$ \\ 
5 (V347 Per)\\ 
5.1,5.2,5.3 & 04 30 49.7 & +34 57 28 & V347 Per? & no & 3 linear flows \\
\noalign{\smallskip}\hline                                       
\end{tabular}

\begin{flushleft}
\textit{Notes}: \\
$^{(a)}$  Coordinates of the heading knot. \\

\end{flushleft}

\end{table*}

\section{slitless observations}

A search for H$\alpha$ emission-line stars in the vicinity of LkH$\alpha$ 101 was previously carried out by \citet{H-A-D}. In total, 35 H$\alpha$ emission-line stars were identified. Of these, 19 were discovered from slitless observations obtained with the UH 2.2-m telescope, while the remaining 16 were subsequently identified from Gemini Multi-Object Spectrograph (GMOS)
data. These objects constitute a compact cluster of pre-main-sequence (PMS) stars associated with NGC 1579 and are concentrated within an area of approximately 10\arcmin.

\begin{figure}
  \centering
  \includegraphics[width=300pt]{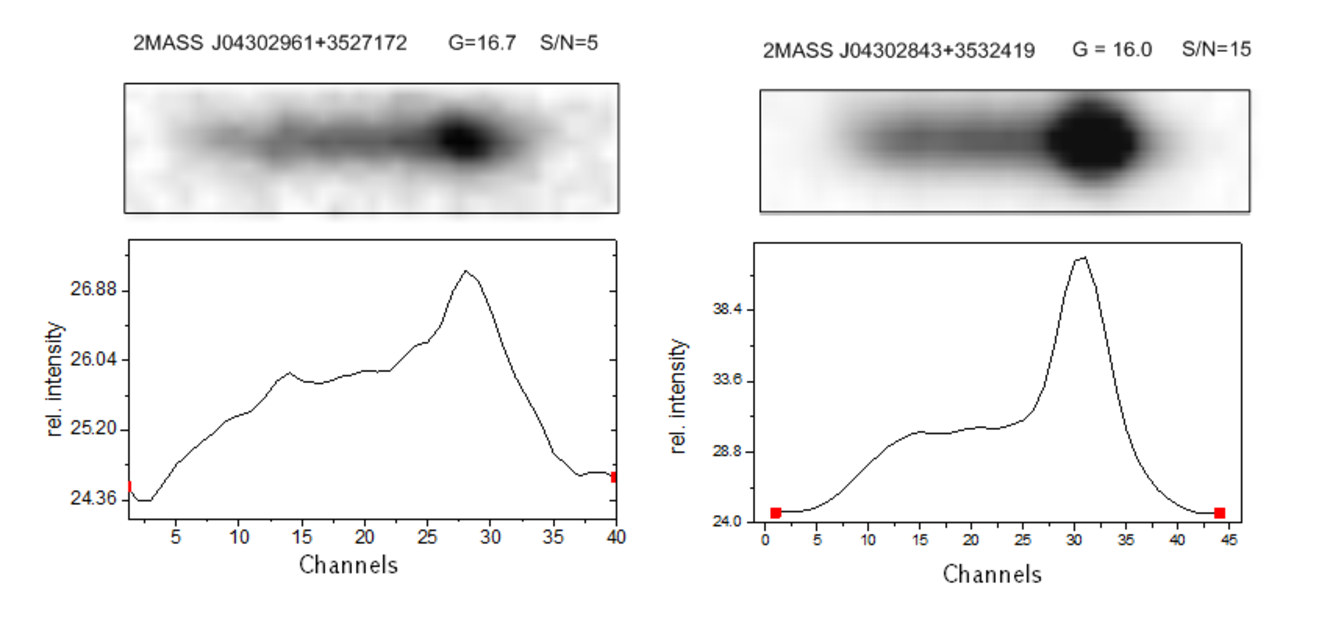}
   \caption{Two spectra of newly found H$\alpha$ emission stars, extracted from the slitless image, and their one-dimensional cuts along dispersion. The S/N ratio are estimated for continuum. Even more faint emission stars still can be found in our image, even if the neighboring continuum is not detected, only by the appearance of H$\alpha$ emission.}
   \label{SLobs}
\end{figure}

As mentioned above, we also carried out slitless spectroscopic observations in this region. Our survey covered a larger field than that of \citet{H-A-D}, but was shifted farther to the north. Consequently, the objects reported by these authors were located outside our observed field. We identified 12 previously unknown H$\alpha$ emission-line stars, which are listed in Table~\ref{hastars}, where we give their identifications by 2MASS survey as well as their G magnitudes and parallaxes (with standard errors) from Gaia DR3 release.
On the Fig.\ref{SLobs} as an example we show two of our slitless spectra with various S/N ratio and their cuts along the dispersion. 

\begin{figure}
  \centering
  \includegraphics[width=200pt]{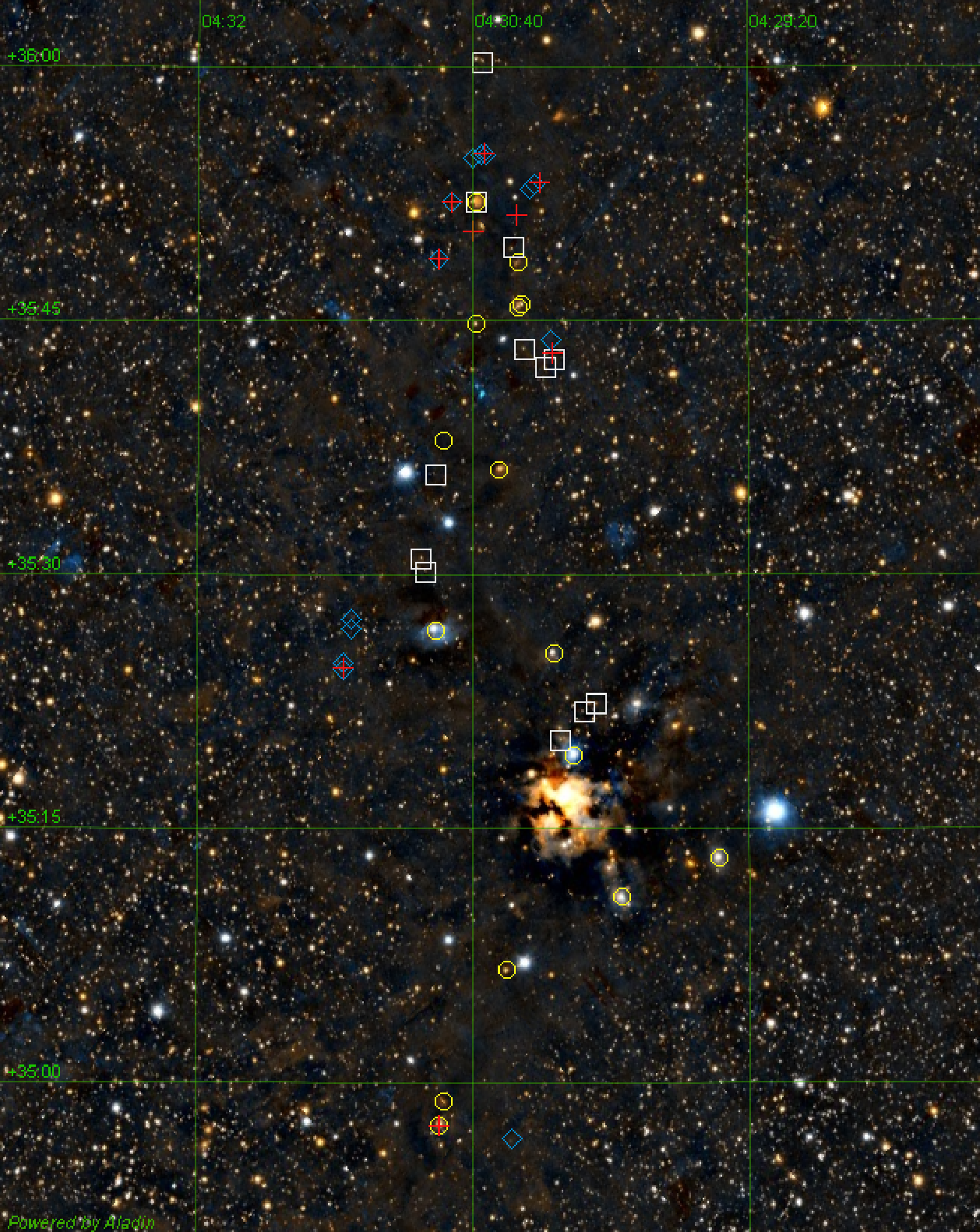}
   \caption{Field to the north from NGC 1579. Emission stars are marked with white squares, nebulous stars - by yellow circles, HH objects - by blue diamonds, and candidate MHO - by red crosses.}
   \label{Hastars}
\end{figure}

Assuming 532$\pm$28 pc for the distance of AMC \citep{Omodaka2020}  we can consider several stars from Table~\ref{hastars} (marked by bold script) as likely members of this aggregate, as indicated by their parallaxes (although there could be certain ambiguities: see notes to the  Table~\ref{hastars} and the discussion in sec.5). Some other emission stars probably constitute its close foreground. Moreover, the  spatial distribution of emission stars on the sky closely follows the northern dark filament of the cloud (Fig.~\ref{Hastars}). Most of these objects are probably in an early stage of stellar evolution. However, spectral classifications are not yet available for the majority of them.

\begin{table*}
\caption{List of H$\alpha$ emission stars discovered by slitless observations}
\label{hastars}
\centering
\setlength{\tabcolsep}{2pt}
{\setstretch{1.5}
\small
\begin{tabular}{l c c l l l }
\hline\noalign{\smallskip}
Object (2MASS ...) & RA (2000.0) & Dec (2000.0) & Other identifiers, sp.type & G(mag)& Plx,   err. (mas)   \\
\hline\noalign{\smallskip}
\textbf{J04300784+3521559} & 04 30 07.8 & +35 21 55.9 & LkHa101CXO J043007.87+352156.0 (M1.0)& 17.81 & 1.9465$\pm$0.1419          \\
 \textbf{J04300424+3522238} & 04 30 04.2 & +35 22 23.9 & LkHa101CXO J043004.29+352224.2 (M0) & 16.54 & 1.8485$\pm$0.0742  \\
 \textbf{J04301473+3520142} & 04 30 14.7 & +35 20 14.3 & LkHa101SST J043014.73+352014.3 & 17.70 & 1.925$\pm$0.1386 \\
 J04305501+3530561 & 04 30 55.0 & +35 30 56.2 & Gaia DR3 174872273516535040 & 17.88 & 2.2599$\pm$0.1743      \\
 \textbf{J04305391+3530109} & 04 30 53.9 & +35 30 11.0 &    Gaia DR3 174872032998363136 & 18.28 & 1.877$\pm0.1735       $ \\
 J04302504+3543180 & 04 30 25.0 & +35 43 18.0 & Gaia DR3 174890106221165184 & 17.44 & 1.6200$\pm$0.1642 \\
 J04301898+3542119 & 04 30 19.0 & +35 42 11.9 &    Gaia DR3 174890041797931904 & 17.91 & not measured \\
 \textbf{J04301628+3542430}$^a$ & 04 30 16.3 & +35 42 42.9 &    Gaia DR3 174890247956361984 & 18.77 & 1.9564$\pm$0.9244 \\
 \textbf{J04302844+3549176} & 04 30 28.4 & +35 49 17.7 &    Gaia DR3 177894556103437952 & 17.31 & 1.9573$\pm$0.1145 \\
 J04303932+3552006$^b$ & 04 30 39.3 & +35 52 00.6 & Gaia DR3 177895110155007360 & 16.25 & 2.0804$\pm$0.0943 \\
 J04303739+3600180 & 04 30 37.4 & +36 00 18.1 &    Gaia DR3 177902905519904384 & 17.40 & 1.6952$\pm$0.2220  \\
 J04305097+3535548 & 04 30 51.0 & +35 35 54.8 & Gaia DR3 174884780461408896 & 19.13 & 2.4982$\pm$0.3544  \\

\noalign{\smallskip}\hline
\end{tabular}
}
\begin{flushleft}
\textit{Notes}: \\
$^{(a)}$ Can be not a member in view of a highly uncertain measurement of parallax (RUWE=4.22)  \\
$^{(b)}$ Also can be a member of AMC; see discussion in sec.5 \\

\end{flushleft}
%\hline 
\end{table*}

\section{Nebulous stars}

Decades of studies of young stellar objects in dark clouds have shown that YSOs associated with small reflection nebulae are, on average, at an earlier evolutionary stage than those lacking such nebulosity. In their comprehensive study of the LkH$\alpha$ 101 cluster, \citet{H-A-D} described six bright nebulous stars, in addition to LkH$\alpha$ 101 itself, illuminating compact reflection nebulae in the vicinity of NGC 1579.

However, even a preliminary inspection of the deeper Pan-STARRS images reveals that this region contains a much larger population of stars associated with small reflection nebulae, particularly within the extended dark filament stretching north of LkH$\alpha$ 101. As described in the previous sections, several of these nebulous stars are associated with ionized and molecular outflows. Others show no evidence of outflow activity, although such activity may be revealed by future observations.

To extend the work of \citet{H-A-D}, we compiled a list of all objects in the vicinity of LkH$\alpha$ 101 that are associated with optical reflection nebulae. The resulting catalogue, which includes only nebulae detected at optical wavelengths, is presented in Table~\ref{nebs}, while the spatial distribution of these objects is shown in Fig.~\ref{Hastars}, along with H$\alpha$ emission stars, HH objects and candidate MHO.

\begin{table*}
\caption{List of nebulous stars around the LkH$\alpha$101}       
\label{nebs}     
\centering
\scriptsize
{\setstretch{1.2}        
\begin{tabular}{l c c p{4.5 cm} l}          
\hline\noalign{\smallskip}                        
 Name & RA (2000.0) & Dec (2000.0) & Descr., other names, sp.type & Plx, err (mas)   \\
   & \ \ h \ m \ s & \ \ \degr\ \ \arcmin\ \ \arcsec\ \\    
\hline\noalign{\smallskip}

V347 Per & 04 30 49.70 & +34 57 27.13 & IRAS 04275+3451, M5:$^b$  (Fig.\ref{south})  & 1.8495$\pm$0.0306\\
IRAS 04275+3452 & 04 30 48.62 & +34 58 53.49 & Gaia DR3 173304408591153664 \newline(Fig.\ref{south})  & 1.9356$\pm0.103$ \\
Gaia DR3 173356223076606464 & 04 30 30.14 & +35 06 39.20 &  Double star (3.5\arcsec) inside \newline common nebula. & 1.7567$\pm0.0838$ \\

Gaia DR3 173360311885473152 & 04 29 57.01 &+35 10 58.63 & Star B, B7$^a$. \newline Illuminates part of  NGC 1579 & 1.8935$\pm0.0257$ \\
2MASS J04292862+3513189 & 04 29 28.63 & +35 13 18.98 & Star A, B9$^a$.  Illuminates part of  NGC 1579 & 1.9066$\pm0.0256$ \\ 
HD 279899 & 04 30 10.88 & +35 19 22.62 &Star E, B4$^a$. \newline Illuminates part of  NGC 1579 & 1.8756$\pm0.0187$ \\
2MASS J04301644+3525217 & 04 30 16.43 & +35 25 21.80 & Star G$^a$. Illuminates faint RN & 1.8883$\pm0.0157$\\
2MASS J04305104+3526453 & 04 30 51.04 & +35 26 45.38 & Star H, B8$^a$. Illuminates \newline conspicuous  RN IC 2067 (see Fig.\ref{center})  & 1.8843$\pm0.0197$\\
IRAS 04272+3529 & 04 30 32.36 & +35 36 13.29 & Star with diffuse nebula \newline near edge of dark cloud. K7e:$^c$ & 1.535$\pm0.1365$  \\
IRAS 04275+3531 & 04 30 48.56 & +35 37 53.46 &  Tiny fan-shaped IR nebula. See sec. 3.3 and refs there. \\
2MASS J04303911+3544497 & 04 30 39.11 & +35 44 49.85 &  Faint nebula visible only in \newline optical range, M7:$^b$ & 1.2596$\pm0.1135$ \\
2MASS J04302705+3545505 & 04 30 27.05 & +35 45 50.47 &  Bipolar nebula with wide cones. \newline See sec. 3.2.3 \\
2MASS J04302615+3545578 & 04 30 26.14 & +35 45 57.87 &  Near the previous one, \newline amorphous. M1.4$^d$. See sec. 3.2.3  & 2.0116$\pm0.0673$ \\
 anon & 04 30 26.83 & +35 48 26.11 &  Very faint star with bipolar \newline nebula. Not visible in IR. \newline (see Fig.\ref{HH1}, SW corner)  \\
2MASS J04303756+3550320 & 04 30 37.56 & +35 50 32.09 &  Star with conspicuous RN, which contains separate knots (however, without embedded stars) and is visible in near IR. M5:$^b$. An IR source with bipolar jet is located nearby. See sec. 3.2.1 and  Fig.\ref{HH1}. \\
2MASS J04303932+3552006 & 04 30 39.34 & +35:52:00.82 & Close to  2MASS J04303916+3552038   star in the nebula, visible also in IR. See sec.3.2.1, sec.5 and  Fig.\ref{HH1}. & 2.0804$\pm$0.0943$^e$\\ 
2MASS J04303916+3552038 & 04 30 39.17 & +35 52 04.14 & Close to  2MASS J04303932+3552006  & 1.8938$\pm0.0903$ \\
IRAS 04269+3550 & 04 30 15.75 & +35 56 57.89  &  Bright conical nebula in mid-IR, while both star (also 2MASS J04301575+3556578) and nebula are very faint in optical range. \newline See sec. 3.2.2 and  Fig.\ref{HH2}.\\
    
\noalign{\smallskip}\hline                                       
\end{tabular}
}
\begin{flushleft}
\textit{Notes}: \\
$^{(a)}$  Star from the list of \cite{H-A-D}. \\
$^{(b)}$ Estimate of sp.type from \cite{Zhang2023} \\
$^{(c)}$ Estimate of sp.type from \cite{cieza} \\
$^{(d)}$ Estimate of sp.type from \cite{birky} \\
$^{(e)}$ Parallax can be overestimated (see sec.5 for discussion)

\end{flushleft}
%\hline 
\end{table*}

It should be noted that many of the spectral types listed in this table were derived using automated classification methods and should therefore be regarded as preliminary. The Gaia parallaxes indicate that most of these stars are indeed members of the AMC, whereas only two objects with measured parallaxes are located beyond the cloud.

One particularly interesting case is the pair 2MASS J04303932+3552006 and 2MASS J04303916+3552038, described in sec. 3.2.1. Although they appear as a close optical double at the centre of a common reflection nebula, their Gaia distances differ for about 40 pc (481 and 528 pc accordingly), while measurement errors are about $\pm$ 22-25 pc. Thus, it seems, that these stars
probably  constitute only an optical pair, and   2MASS J04303932+3552006 represents an object of close foreground of AMC. However, this problem is more complicated than it seems. The careful examination of the PanSTARRS image reveals that both  2MASS J04303932+3552006 and 2MASS J04303916+3552038 stars not only are enveloped by a reflection nebula (which has some ring-like structures near 2MASS J04303932+3552006), but this nebula shows local brightening near both stars. This suggests that these stars both are connected with nebula and should belong to AMC cloud; the Gaia parallax of  2MASS J04303932+3552006 may be erroneous since its RUWE (Re-normalised Unit Weight Error ) is quite large - 1.384 (possibly due to the influence of the nebula). Besides, as stated in sec.4, 2MASS J04303932+3552006 has H$\alpha$ emission and is a good candidate to be a progenitor for HH 1276 and MHO 1.4 (see sec. 3.2.1). This star definitely deserves further studies.

\section{Conclusion}

As noted above, the Auriga-California Molecular Cloud (AMC) exhibits a moderate level of star formation activity, as indicated by the population of YSOs identified in the \textit{Spitzer} Gould Belt Survey and by \textit{Herschel} observations \citep{Harvey2013,Broekhoven}. Virtually all evolutionary classes, from Class 0 to Class II, are concentrated along the high-column-density filamentary structure identified by \citet{Harvey2013} (their Fig.~8). A comparison with our Fig.~\ref{Hastars} shows that the newly discovered HH objects and MHOs are distributed along the same filament. This provides additional evidence that star formation is currently concentrated within this dense structure.

On the other hand, among the H$\alpha$ emission-line stars identified in our slitless survey, only one object, 2MASS J04303932+3552006, can tentatively be considered the driving source of a collimated outflow (see discussion in sec.5). This result suggests that the majority of the outflow-driving sources remain too deeply embedded to be detected in optical slitless spectroscopy.

Compared with the Orion Molecular Cloud, the number of HH objects and outflows identified in the AMC is relatively small, although their projected lengths are comparable. Furthermore, nearly all of the optical outflows have molecular hydrogen counterparts, frequently forming well-defined MHO flows. This strongly suggests that the outflow systems are deeply embedded within the molecular cloud and that only their least obscured optical components are detectable.

Taken together, our results indicate that the driving sources of the collimated outflows observed as HH objects and MHOs in the AMC are predominantly deeply embedded Class 0 and Class I young stellar objects.
Follow-up near-infrared spectroscopy and high-resolution molecular-line observations will be essential for identifying the driving sources of the newly discovered outflows and for investigating their physical properties.

\section*{Acknowledgements}

We thank Prof. B. Reipurth for assigning the numbers to the newly discovered HH objects. We also thank the referee for many very useful comments which allowed to greatly improve this paper.

This research has made extensive use of \textit{Aladin} sky atlas, \textit{VizieR} catalogue access tool,  \textit{VizieR}
photometry tool and \textit{SIMBAD} database, which are developed and
operated at CDS, Strasbourg Observatory, France.  This work has made use of data from the European Space Agency (ESA) space mission Gaia, processed by the Gaia Data Processing and Analysis Consortium (DPAC). Funding for the DPAC is provided by national institutions, in particular the institutions participating in the Gaia MultiLateral Agreement (MLA). The Gaia mission website is \textit{https://www.cosmos.esa.int/gaia}. The Gaia archive website is \textit{https://archives.esac.esa.int/gaia}. This publication makes use of data products from the Two Micron All Sky Survey (2MASS), which is a joint project of the University of Massachusetts and the Infrared Processing and Analysis Center/California Institute of Technology, funded by the National Aeronautics and Space Administration and the National Science Foundation. This publication also makes use of data products from the Wide-field Infrared Survey Explorer (WISE), which is a joint project of the University of California, Los Angeles, and the Jet Propulsion Laboratory/California Institute of Technology, funded by the National Aeronautics and Space Administration. The images from Spitzer Enhanced Imaging Products program were used via NASA/IPAC
Infrared Science Archive \textit{https://www.ipac.caltech.edu/doi/irsa/10.26131/IRSA3}

%%%%%%%%%%%%%%%%%%%%%%%%%%%%%%%%%%%%%%%%%%%%%%%%%%
\section*{Data Availability}

The data underlying this article will be shared on reasonable request to the corresponding author. Tables \ref{MHOs}, \ref{hastars} and \ref{nebs} will be deposited in electronic form in \textit{VizieR} database.

%%%%%%%%%%%%%%%%%%%%%%%%%%%%%%%%%%%%%%%%%%%%%%%%%%

\end{document}